# AC Losses of Copper Stabilized Multifilament YBCO Coated Conductors

G. A. Levin, J. Murphy, T. J. Haugan, J. Šouc, J. Kováč, and P. Kováč

*Abstract* — We report the data on magnetization losses and critical current of multifilament copper stabilized coated conductors. Eight centimeters long samples of copper stabilized YBa$_2$Cu$_3$O$_{7-x}$ (YBCO) coated conductors manufactured commercially were subdivided into superconducting filaments by near-IR laser micromachining. The width of the superconducting stripes was varied from 0.2 mm to 0.04 mm. Some of the samples were striated leaving superconducting bridges for current sharing between the filaments. The AC losses were measured at different sweep rates of the magnetic field up to 14 T/s. We will present the results for the hysteresis and coupling losses and discuss the means to reduce the coupling loss by changing the processing parameters of micromachining and by post-ablation treatment.

## I. INTRODUCTION

Over the years there has been a significant effort by several research groups, as well as by the manufacturers of coated conductors, to implement striation as a means to reduce the AC losses in the future power cables. By far an incomplete list of such publications includes references [1-9]. Another important small scale application of striated coated conductors is integration on the same substrate of multiple mini-current leads to feed cryo-electronics [10,11]. For both of these purposes (AC power cables and current leads) the stabilization of the superconducting filaments by copper or silver layer of substantial thickness is necessary to avoid a catastrophic (one that results in irreversible damage) quench. The individual stripes must be well insulated from each other in order to avoid the coupling losses, which may defeat the purpose of striation, and to avoid the cross-talk in current leads. Most of the early experiments with striated coated conductors were carried out, as a proof of concept, on non-stabilized samples with only a thin (1-3 μm) silver layer on top of YBCO film. Later, some groups went farther, and carried out the striation of coated conductors with a stabilizer applied [8,12].

Here we present the results of measurements of the critical currents and AC losses in fully stabilized and laser striated coated conductors. The width of the individual stripe was varied from 0.2 to 0.04 mm. The thickness of copper stabilizer striated along with the superconducting layer was 20 μm. We have determined the contributions of the hysteresis, coupling and eddy current loss to the total. One of the conclusions we draw from this sequence of experiments is that fine striation, while decreasing the overall critical current, increases the pinning strength which translates into less steep fall of the critical current in magnetic field. The striation leads to substantial reduction of AC loss relative to the non-striated sample. However, the ac loss per unit of critical current does not decrease below the level which can be achieved in the sample with 1 mm wide stripes.

## II. EXPERIMENTAL DETAILS

### A. Sample preparation

Laser micromachining of coated conductors is a method that has been tested thus far more extensively than its potential alternatives (photolithography, etc.). We have used a diode-pumped solid-state femtosecond laser at 1033 nm wave length to cut grooves in the 12 mm wide samples of copper stabilized coated conductors produced by SuperPower Inc. The 20 μm thick copper stabilizer layer was wrapped around the whole tape (surround copper stabilizer), but only one side of the stabilizer, which is in contact with the superconducting film, was striated. The back side of the stabilizer was not striated and had served as an additional source of eddy current losses. In some samples superconducting bridges were left in order to allow the current exchange between the individual stripes as a measure of increasing reliability of the overall conductor [3-5,12,13].

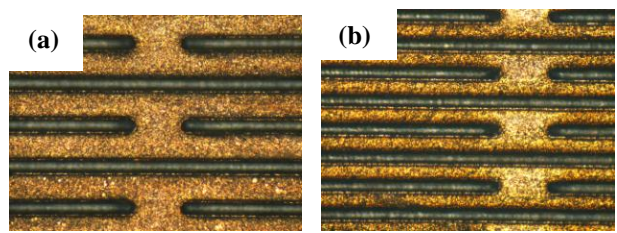

**Figure 1(a,b)** Striated samples with the laser cut grooves (shown at 10x magnification). The width of the individual superconducting stripes is 0.07 mm in Fig. 1(a) and 0.04 mm in Fig. 1 (b). Superconducting bridges facilitating current transfer between the individual stripes were left in some samples.

### B. Critical current measurements

Multiple, parallel 8 cm long grooves were cut in the central section of 12 cm long tapes leaving sufficient unstriated margins at both ends in order to measure the critical current $I_c$ using a standard 4-point method. The voltage criterion for determining $I_c$ was chosen as 1 μV/cm. The measurements

Manuscript received October 9, 2012. This work was supported by the U.S. Air Force Office of Scientific Research and its European Office of Aerospace Research and Development.

G. A. Levin is with UES, Inc. Dayton, OH 45432 USA (937-255- 5630, George.levin.ctr@wpafb.af.mil).
J. Murphy is with the University of Dayton Research Institute, Dayton OH 45469, USA.
T. J. Haugan is with the Aerospace Systems Directorate, Air Force Research Laboratory, Wright-Patterson AFB OH 45433, USA
J. Šouc, J. Kováč, and P. Kováč are with the Institute of Electrical Engineering, Slovak Academy of Sciences, 841 04 Bratislava, Slovakia.

were carried out for both directions of the applied current and yield very close results. The magnetic field $B$ was applied perpendicular to the wide face of the tape. The temperature was kept at 77 K by immersing the samples in liquid nitrogen.

*C. AC loss measurements.*

After the critical current was determined, the unstriated ends were cut resulting in 8 cm long fully striated samples for measuring the AC losses. The AC loss measurement method is based on the measurement of a part of the power supplied by the AC source to the AC solenoid generating the magnetic field. The measurement system consists of two identical ac copper solenoids connected in series and two measurement pick-up coils wound in parallel with magnets windings. During the measurements the complete system is immersed in liquid nitrogen. The superconducting sample is located inside one of the solenoids, with the other solenoid and a pick-up coil serving as compensation. More details about the calibration free method for AC loss measurement are described in Ref. [14].

## III. RESULTS

*A. Critical current*

Figure 2(a) shows the critical current of several striated samples as a function of magnetic field. As expected, the greater the number of stripes, the smaller the critical current due to removal of successively larger amount of superconducting material. The width of the non-superconducting grooves is approximately 30 μm. In Fig. 2(b) the critical currents of all samples are normalized to their value at zero field. Here we see that the finely striated samples, especially the ones divided into 0.04 or 0.07 mm wide stripes retain the value of the critical current better than the non-striated or more coarsely striated conductors. For example, the critical current of the non-striated sample falls to 50% of its zero field value in field of 150 mT. The sample with 40 μm wide stripes has the critical current equal to 64% of its zero field value in the same field.

The magnetic self-field pattern slightly depends on the number of stripes. This may affect the critical current and it does, but insignificantly. The data in Fig. 2 show the values of $I_c$ averaged over two opposite directions of current with the same direction of the external applied field. Thus, the first order effect of self-field on $I_c$ cancels out. This leaves us with a conclusion that the striation increases the pinning strength. Of course, nobody suggests striation as a means to improve the current-carrying capacity of the coated conductors. These results, however, lend additional support to experimental findings and theory considerations advanced in Ref. [15] that the edge-barrier pinning leads to enhanced critical current in field. Although 0.04 or 0.07 mm wide stripes seem too wide to have such a pronounced effect, other types of damage caused by laser ablation may have a beneficial side-effect increasing the number and strength of the pinning centers.

*B. AC losses.*

Striated conductors like the ones shown in Fig. 1 have three main components of AC loss. One component is the hysteresis loss in superconducting material. In magnetic field above the penetration threshold [16]

$$P_h \approx I_c W_n B f . \quad (1)$$

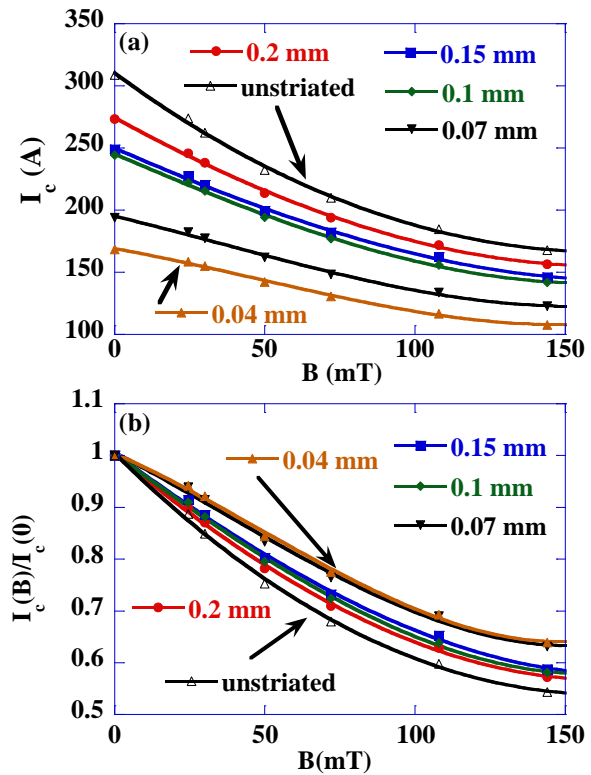

Figure 2(a,b) (a) Measured critical current vs applied magnetic field for several striated and one non-striated sample. The width of the stripes is indicated. (b) The same data as in Fig. 2(a), but the currents are normalized to their values in zero applied field. Arrows indicate the data for non-striated sample and the one with the maximum number of stripes, each 0.04 mm wide.

Here $P_h$ is the hysteresis loss per unit length, $W_n$ is the width of an individual stripe and $Bf$ is the sweep rate, where $B$ is the peak value of the alternating field and $f$ is its frequency. Sweep rate characterizes the average value of the time derivative $|\dot{B}| \propto Bf$. The eddy current loss in these samples is predominantly takes place in the back side of the stabilizer, which is left unstriated. In the full penetration regime the eddy current loss per unit length in a rectangular tape is given by [17,18]

$$P_{e-c} \approx \frac{\pi^2}{6} \frac{dW^3}{\rho} (Bf)^2 . \quad (2)$$

Here $d$ is the thickness of the metal stripe, $W$ is its width, and $\rho$ is the resistivity. The third component of ac loss originates from the remaining resistive coupling between the superconducting stripes and is also proportional to the square of the sweep rate. By analogy with Eq. (2) we can write down a phenomenological expression for the coupling losses [4,19-22] as

$$P_c \approx \frac{\pi^2}{6} \frac{L^2 W}{R_{eff}} (Bf)^2 . \quad (3)$$

Here the effective resistance $R_{eff}$ characterizes the coupling strength. The total ac loss [W/m]

$$P = P_h + P_c + P_{e-c} \quad (4)$$

can be presented in the form of the power loss per unit length, per unit of the sweep rate, per unit of the critical current as a function of the sweep rate

$$\frac{P}{I_c Bf} = \lambda_1 + \lambda_2 Bf \ ; \quad \left[\frac{J}{mAT}\right] \equiv [m]. \quad (5)$$

This allows us to segregate the hysteresis loss from the eddy current and coupling losses as the intercept $\lambda_1$ and the slope $\lambda_2$. If the striation is carried out perfectly, so that there is no superconducting links between the stripes, and the critical current is not degraded beyond the loss of superconducting material, the value of the intercept will be close to the width of the stripes, see Eq. (1), as indeed was the case in the striated samples without stabilizers [4,13,20,21]. However, if the intercept $\lambda_1$ is substantially greater than $W_n$, it may be an indication of remaining superconducting connections between the stripes, or a more serious degradation of the critical current (by overheating and loss of oxygen, for example, during striation).

*C. Control experiment.*

As a control experiment we used a coated conductor covered only with a thin silver layer. The 12 mm wide sample was striated into 0.5 mm wide stripes, its critical current and AC losses were measured as described above. Figure 3 shows the power loss presented in the form of Eq. (5) as the power loss per unit length, per unit of the sweep rate, per unit of the critical current. The linear fit gives us the values of the intercept and the slope. The intercept exactly matches the width of the individual stripes indicating a perfect striation (absence of superconducting connections between the stripes). The slope indicates substantial coupling loss due to normal metal coupling between the stripes. The value of the slope shown in Fig. 3 is that obtained by a fit to 72 Hz data. The results shown in Fig. 3 are similar to those obtained earlier on similar samples [4,20].

The value of the slope determines the sum of the coupling and eddy current losses. The contribution of the eddy current losses to the slope is given by Eq. (2):

$$\lambda_2^{e-c} = \frac{\pi^2}{6} \frac{dW^3}{\rho I_c} \ . \quad (6)$$

In this sample the eddy current losses are determined by the 50 μm thick Hastelloy substrate with the resistivity $\rho \approx 120 \times 10^{-6} \Omega cm$. Taking W=12 mm and $I_c$=217.6 A, we get

$$\lambda_2^{e-c} \approx 5.4 \times 10^{-4} mm \cdot s \cdot T^{-1}. \quad (7)$$

Comparing this with the total value of $\lambda_2$= *0.16 mm s $T^{-1}$* shown in Fig. 3 we conclude that the main contribution to the loss with quadratic dependence on the sweep rate is the coupling loss originated from the metal connections between the stripes. When the striated sample was later annealed in oxygen, the coupling component of the total loss disappeared, because metal oxides that has formed in the grooves insulated

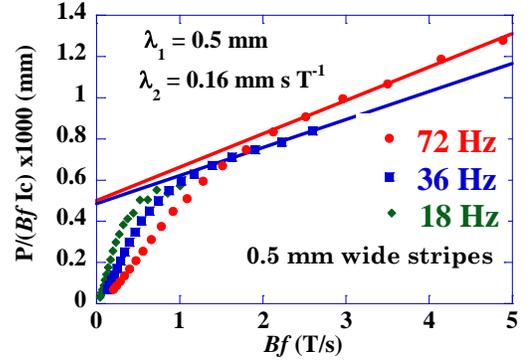

Figure 3. Power loss per unit length, per unit of the sweep rate, per unit of the critical current, Eq. (5). The critical current $I_c$ = 217.6 A. The intercept $\lambda_1$=0.5 mm and the slope $\lambda_2$ are the measures of the hysteresis loss and coupling loss, respectively. The straight lines correspond to the linear fit of 72 and 36 Hz data.

the superconducting stripes from each other [21,22].

*D. Losses in striated and copper stabilized samples.*

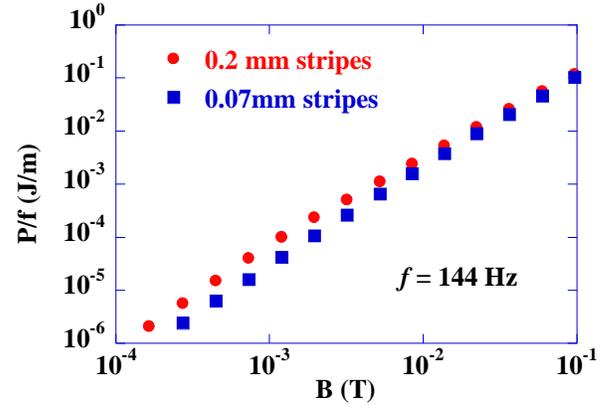

Figure 4. AC loss per cycle as a function of the applied field amplitude. Shown are the data for two samples, one with 0.2 mm wide stripes and the other with 0.07 mm wide stripes. In both cases the frequency is 144 Hz. The grooves segregating the stripes were cut through the 20 μm thick stabilizer and YBCO film.

Figure 4 presents the loss per cycle in two copper stabilized samples with 0.2 mm and 0.07 mm wide stripes respectively. The loss decreases with the number of stripes, but it is difficult to judge from this traditional presentation whether the reduction of losses is the result of degradation of critical current, or a genuine improvement.

Figure 5 shows the data for the same sample as in Fig. 4 (0.2 mm wide stripes) presented in the form of Eq. (5) for two different frequencies of the applied magnetic field. The data for both frequencies strongly overlap, confirming the universality of the description given by Eq. (5). The value of the critical current was measured in zero applied field, Fig. 2(a). The sample is divided into approximately 50 stripes, 0.2 mm wide each with approximately 0.03 mm wide grooves segregating them. The intercept $\lambda_1$=1.2 mm is significantly greater than the expected value $\lambda_1$=0.2 mm. There are two possible reasons for that. One possibility is that there might still be some superconducting coupling between the stripes.

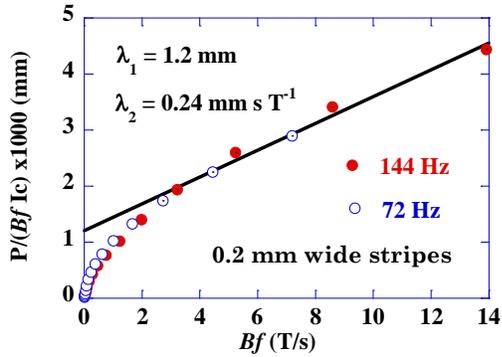

Figure 5. AC power losses in the form of Eq. (5) in the sample with 0.2 mm wide stripes for two different frequencies 77 and 144 Hz. The grooves segregating the stripes were cut through the 20 μm thick stabilizer and YBCO film. The critical current $I_c$=273 A. The line is a linear fit to the data which includes both frequencies.

In the absence of such connections, the intercept would be close to 0.2 mm, equal to the width of the stripe (see Fig. 3). The hysteresis loss per unit of critical current in this 50 stripe sample is equivalent to that in a perfectly striated sample with 10 stripes, each 1.2 mm wide. If there are superconducting connections between the stripes they are likely formed by a random network of microscopically small superconducting links that have survived the laser ablation.

Another possibility is that the laser ablation of the superconducting film through the relatively thick stabilizer not only removes some amount of superconducting material, as in the case of the control sample, but does substantial additional damage to the remaining superconducting film. As the result, the hysteresis loss *per unit of critical current*, which is what Eq. (5) defines, is greater than expected. This consideration underscores the importance of evaluating AC losses relative to the current-carrying capacity of the conductor.

The eddy current loss in this sample is dominated by the copper stabilizer on the back side of the conductor. It is a non-striated strip of copper 20 μm thick, 12 mm wide with the nominal resistivity at 77 K

$$\rho \approx 3 \times 10^{-7} \Omega cm.$$

The critical current of this sample in zero field, Fig. 2(a), is 273 A. Thus, the contribution of the eddy current loss in the back side of the surround stabilizer to the value of the slope given by Eq. (6) is

$$\lambda_2^{e-c} \approx 0.07 mm \cdot s \cdot T^{-1}. \qquad (8)$$

In the full penetration regime one can expect that the value of $\lambda_2$=0.24 mm s $T^{-1}$ is the sum of $\lambda_2^{e-c} + \lambda_c$, where $\lambda_c$ obtained from Eq. (3) determines the coupling loss. Thus, we see that the eddy current loss in these samples is about one half of the coupling losses. From the value of $\lambda_c$=0.17 mm s$T^{-1}$ we can estimates the effective coupling resistance $R_{eff}$ ~0.67 mΩ. The value of $\lambda_c$ which determines the coupling loss is very similar to that in the control sample, Fig. 3, and to that in the previously investigated samples of comparable length and width [4,20], but without copper stabilization. This indicates that the resistive coupling between the superconducting filaments is due to substrate metal (Hastelloy), rather than copper.

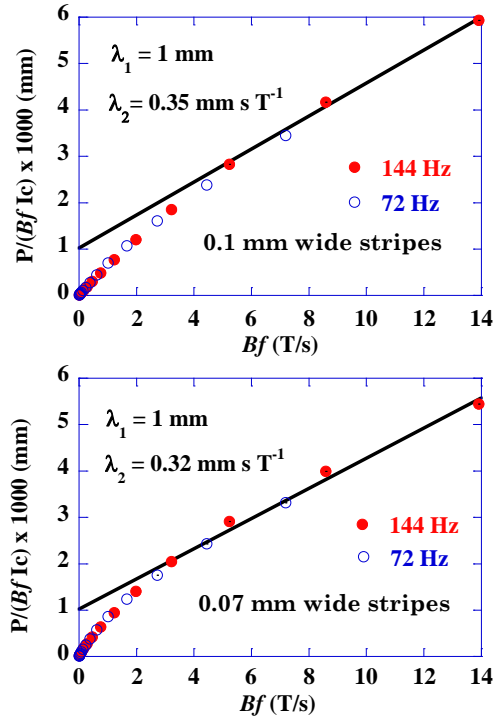

Figure 6. AC power losses in the form of Eq. (5) in another two copper stabilized samples divided into 0.1 mm and 0.07 mm wide stripes for two different frequencies 77 and 144 Hz. The grooves segregating the stripes were cut through the 20 μm thick stabilizer and YBCO film. The critical currents are 244 A and 194 A respectively. The straight lines are a linear fit to the data which includes both frequencies.

Figure 6 shows the data for two samples with finer stripes, 0.1 and 0.07 mm wide, respectively. The AC loss decreases with increasing number of stripes, but the effect seems to be entirely due to removal of superconducting material. Normalized to the critical current, as shown in Fig. 6, the hysteresis loss and the coupling and eddy current losses practically remain the same.

## IV. CONCLUSION

Previously, two approaches to making striated and stabilized coated conductors were outlined [22]. One was a "brute force" approach which is documented in this report. As it turns out, in these samples the total loss per unit of the critical current is about the same as in striated samples with 1 mm wide stripes. Perhaps, more promising results can be achieved by implementing the second route and striating first a silver coated conductor with subsequent oxygen annealing. The formation of insulating metal oxides on the surface of the grooves may allow selective electro-deposition of copper only on the surface of silver coated superconducting stripes. The first successful demonstration of this approach, with mechanically scribed grooves, is reported in Ref. [23].

## ACKNOWLEDGMENT

This work was supported in part by the AFOSR.

REFERENCES

[1] N. Amemiya, S. Kasai, K.Yoda, Z. Jiang, G. A. Levin, P. N. Barnes, and C. E. Oberly, AC Loss reduction of YBCO coated conductors by multifilamentary structure," *Superconductor Science Technology* vol. 17, pp. 1464-1471 (2004).

[2] M.D. Sumption, E.W. Collings, and P.N. Barnes, "AC loss in striped (filamentary) YBCO coated conductors leading to designs for high frequencies and field-sweep amplitudes" *Supercond Sci. Technol.*, 18, pp. 122-134 (2005).

[3] M. Majoros, B. A. Glowacki, A. M. Campbell, G. A. Levin, P. N. Barnes, and M. Polak, "AC Losses in striated YBCO coated conductors," *IEEE Transactions on Applied Superconductors* vol. 15, p. 2819, (2005).

[4] G. A. Levin, P. N. Barnes, N. Amemiya, S. Kasai, K. Yoda, Z. Jiang, and A. Polyanskii, "Magnetization losses in multiply connected $YBa_2Cu_3O_{6+x}$ coated conductors," *Journal of Applied Physics*, vol. 98, p. 113909, (2005)

[5] P. N. Barnes, M. D. Sumption, and G. L. Rhoads, "Review of high power density superconducting generators: Present state and prospects for incorporating YBCO windings," *Cryogenics* vol. 45 (2005) pp. 670–686.

[6] P. N. Barnes, G. A. Levin, C. Varanasi, and M. D. Sumption, "Low AC Loss Structures in YBCO Coated Conductors With Filamentary Current Sharing", *IEEE Trans. Appl. Supercond*. 15, p. 2827 (2005).

[7] V. Selvamanickam et al. "Progress in performance improvements and new research areas for cost reduction of 2G HTS wires," *IEEE Transactions on Applied Superconductors* Vol. 21, pp. 3049 (2011)

[8] S. Terzieva et al. "Investigation of the effect of striated strands on the AC losses of 2G Roebel cables" *Supercond Sci. Technol.*, 24, 145001 (2011).

[9] R. J. Webber, J. Delmas, and B. H. Moeckly, "Ultra-Low Heat Leak YBCO Superconducting Leads for Cryoelectronic Applications", *IEEE Trans. on Appl. Supercond.* 19 (3), p. 999 (2009)

[10] R. J. Webber and J. Delmas, "Operation of YBCO current leads as Bias Lines to Cryocooler Mounted 4 K Superconducting Electronics" *Physics Procedia*, 36, p. 256 (2012)

[11] H-S. Shin et al. "Electro-Mechanical Property Investigation of Striated REBCO Coated Conductor Tapes in Pure Torsion Mode" *IEEE Trans. on Appl. Supercond.* 21 2997 (2011)

[12] G. A. Levin and P. N. Barnes, "Concept of Multiply Connected Superconducting Tapes", *IEEE Trans. Appl. Supercond.* 15, p. 2158 (2005).

[13] G. A. Levin and P. N. Barnes, "*Second Generation Superconducting Wires for Power Applications*" in "Flux Pinning and AC Loss Studies on YBCO Coated Conductors, P. Paranthaman and V. Selvamanickam, Eds. Nova Science, New York (2007) pp. 299-327

[14] Jan Souc, Fedor Gomory and Michal Vojenciak, "Calibration Free Method for Measurement of the AC Magnetization Loss" *Supercond. Sci. Technol.* 18, 592 (2005)

[15] W. A. Jones et al. "Impact of edge-barrier pinning in superconducting thin films", *Appl. Phys. Lett*. 97, 262503 (2010)

[16] E. H. Brandt and M. Indenbom, "Type-II-Superconductor Strip with Current in a Perpendicular magnetic field", *Phys. Rev*. B 48, 12893 (1993)

[17] K.-H. Müller, "AC Power Losses in Flexible thick-film superconducting tapes", Physica C 281, p. 1 (1997)

[18] D. N. Nguyen, P. V. P. S. S. Sastry, and J. Schwartz, "Numerical calculations of the total ac loss of Cu-stabilized $YBa_2Cu_3O_{7-\delta}$ coated conductor with a ferromagnetic substrate" *J. Appl. Phys.*, 101, 053905 (2007)

[19] W. J. Carr Jr. and C. E. Oberly, "Filamentary YBCO conductors for AC applications", *IEEE Trans. Appl. Supercond*. 9, p. 1475 (1999).

[20] G. A. Levin, P. N. Barnes, N. Amemiya, S. Kasai, K.Yoda, and Z. Jiang, "Magnetization losses in multifilament coated superconductors", *Appl. Phys. Lett*. 86, 072509 (2005).

[21] G. A. Levin, P. N. Barnes, J. W. Kell, N. Amemiya, Z. Jiang, K. Yoda, and F. Kimura, "Multifilament $YBa_2Cu_3O_{6+x}$-coated conductors with minimized coupling losses", *Applied Physics Letters*. vol. 89, 012506 (2006)

[22] G. A. Levin, P. N. Barnes, and N. Amemiya, "Low ac Loss Multifilament Coated Conductors", *IEEE Trans. Appl. Supercond.* 17 (2), p. 3148 (2007)

[23] I. Kesgin, G. Majkic, and V. Selvamanickam, "Fully filamentized HTS coated conductor via striation and selective electroplating," *Physica C,* (in print).